\documentclass{JHEP3} 

\usepackage{epsfig}
\usepackage{amsmath, amssymb}
\usepackage{concmath, palatino}
\usepackage{mathrsfs}
\usepackage{epic}
\usepackage{color}

\newcommand\fverb{\setbox\pippobox=\hbox\bgroup\verb}
\newcommand\fverbdo{\egroup\medskip\noindent%
			\fbox{\unhbox\pippobox}\ }
\newcommand\fverbit{\egroup\item[\fbox{\unhbox\pippobox}]}
\newbox\pippobox

\newcommand{\be}{\begin{equation}}
\newcommand{\ee}{\end{equation}}
\newcommand{\ba}{\begin{eqnarray}}
\newcommand{\ea}{\end{eqnarray}}

\newcommand{\la}{\longrightarrow}

\newcommand{\ads}{AdS_5\times S^5}

\newcommand{\ddb}{{\overline{\mathscr D}}}

\newcommand{\N}{\mathcal{N}}

\allowdisplaybreaks

    \newcommand{\beq}{\begin{equation}}
    \newcommand{\eeq}{\end{equation}}
    \newcommand\beqa{\begin{eqnarray}}
    \newcommand\eeqa{\end{eqnarray}}

\title{Wrapping corrections, reciprocity and BFKL beyond the $\mathfrak{sl}(2)$ subsector in $\N$ = 4 SYM}

\author{Matteo Beccaria\\
  Dipartimento di Fisica, Universita' del Salento, 
  Via Arnesano, 73100 Lecce \&\\
  INFN, Sezione di Lecce\\
  E-mail: \email{matteo.beccaria$\bullet$le.infn.it}
}

\author{Guido Macorini\\
  Dipartimento di Fisica, Universita' del Salento, 
  Via Arnesano, 73100 Lecce \&\\
  INFN, Sezione di Lecce\\
  E-mail: \email{guido.macorini$\bullet$le.infn.it}
}

\author{CarloAlberto Ratti\\
  Dipartimento di Fisica, Universita' del Salento, 
  Via Arnesano, 73100 Lecce \&\\
  INFN, Sezione di Lecce\\
  E-mail: \email{carloalberto.ratti$\bullet$le.infn.it}
}

\abstract{
We consider  $\N$ = 4 SYM and a class of spin $N$, length-3, twist operators  beyond the well studied $\mathfrak{sl}(2)$ subsector. They can be identified at one-loop with 
three gluon operators. At strong coupling, they are associated with  spinning strings with two spins in $AdS_{5}$.
We exploit the Y-system to compute the leading weak-coupling four loop wrapping correction to their anomalous dimension. The result is written in closed form as a function of the spin  $N$. We combine the wrapping correction with the known four-loop asymptotic Bethe Ansatz contribution and analyze special limits in the spin $N$. In particular, at large $N$, we prove
that a generalized Gribov-Lipatov reciprocity holds. At negative unphysical spin, we present a simple BFKL-like equation predicting the rightmost leading poles.
}

\begin{document} 

\section{Introduction}
\label{sec:intro}

The study of finite size corrections to states/operators in AdS/CFT correspondence can be addressed 
within the powerful and general approach of the mirror thermodynamic Bethe Ansatz (TBA) developed for the 
$\ads$ superstring in~\cite{TBA1} and deeply tested in \cite{TBA2}, mainly in $\mathfrak{sl}(2)$ closed 
subsector. 
The associated Y-system has been proposed in \cite{GKV} based on symmetry arguments and educated guesses
about the analyticity and asymptotic properties of the Y-functions. Further developments can be found in \cite{FurtherKolya}.

\medskip
Very powerful explicit tests of wrapping corrections can be done in the $\mathfrak{sl}(2)$ sector of $\N=4$ SYM.
The relevant operators are represented by the insertion of $N$ covariant derivatives $\mathcal{D}$ into  the protected half-BPS state $\mbox{Tr} \mathcal{Z}^{L}$  ($\mathcal Z$ being one of the three complex scalars)
\be
\mathbb{O}_{N, L}^{\mathcal Z} = \sum_{s_{1},\dots, s_{L}} c_{s_{1}, \dots, s_{L}}\,\mbox{Tr} \left( \mathcal{D}^{s_1}\mathcal{Z}\cdots\mathcal{D}^{s_L}\mathcal{Z}\right),
\quad  \mathrm{with} \quad N=s_1+\cdots+s_{L}\,. 
\ee
Their anomalous dimensions can be obtained from a non-compact, length-$L$ $\mathfrak{sl}(2)$ spin chain 
with $N$ excitations underlying a factorized two-body scattering. The 
interaction range between scattering particles increases with orders of the coupling constant in perturbation theory.
If it exceeds the length of the spin chain and {\it wraps} around it, the 
S-matrix picture  fails, as no asymptotic region can be defined any longer. 
For twist $L$ operators this effect, delayed by superconformal invariance, starts at order $g^{2L+4}$. 

\medskip
The most advanced available calculations are at five-loop order for $L=2$~\cite{Lukowski:2009ce} and at 
six-loop order for $L=3$~\cite{Velizhanin:2010cm}. The anomalous dimension of $\mathbb{O}_{N, L}^{\mathcal Z}$
is presented in closed form as a function of $N$. This allows to study special limits which have physical significance
and test the TBA framework. In particular, at large $N$, it is found that a generalized Gribov-Lipatov
reciprocity (see  \cite{reciprocity} and the recent review \cite{Beccaria:2010tb}) holds predicting basically half of the expansion in terms of the other half. Also, it is possible to match the predictions of the BFKL
equation~\cite{Lipatov:1976zz} governing the poles around 
unphysical negative values of $N$. Notice that both reciprocity and BFKL are not related to integrability
and have a wider physical meaning. As such, we regard them as  independent checks of integrability 
predictions~\footnote{See \cite{Velizhanin:2011pb} for a recent analysis of the interplay between reciprocity and 
BFKL poles.}.

\medskip
The tests in the $\mathfrak{sl}(2)$ explore a relatively small part of the full $\mathfrak{psu}(2,2|4)$ structure of $\N=4$ SYM.
For this reason, we believe that similar calculations in a larger sector would be very interesting. Of course, for 
general states, there is no reason to expect to be able to find similar closed formulae in any parameter and the 
proposal seems hopeless. A remarkable and rather peculiar exception are the twist operators studied at many loops 
in~\cite{Beccaria:2007pb}. We shall refer to these operators as {\em 3-gluon operators} $\mathbb{O}_{N, 3}^{\mathcal A}$. The reason for this terminology is that at one-loop they have the same form as $\mathfrak{sl}(2)$ operators, 
with the scalar $\mathcal Z$ being replaced by a physical gauge field component. At strong coupling, 
these operators are part of a larger family studied in \cite{Freyhult:2009fc} and dual to spinning string configurations
with two spins in $AdS_{5}$. As discussed in~\cite{Beccaria:2007pb}, it is possible to derive the asymptotic (with no wrapping corrections) anomalous dimensions in closed forms as functions of $N$. Wrapping corrections are expected to 
appear at four loops. Up to this level, reciprocity holds for the asymptotic contribution. The BFKL poles have not been studied yet.

\medskip
Thus, a quite reasonable plan is that of computing the four-loop wrapping corrections exploiting the structural features of the asymptotic contributions in order to determine a closed form for the wrapping as a function of the spin $N$. This is precisely what we shall describe in this paper. As a byproduct, we shall be able to 
test (positively) reciprocity as well as discuss the BFKL poles of the full four-loop result. Our analysis provides
an extension of the well-known results for scalar twist-operators to the full $\mathfrak{psu}(2,2|4)$  states.
Notice indeed, that beyond one-loop the 3-gluon operators have a non trivial mixing with other operators and 
do not belong to a closed subsector.

\medskip
The plan of the paper is the following. In Sec.~(\ref{sec:gluonic}), we recall the details of the 
operators under consideration. In Sec.~(\ref{sec:oneloop}), we present their one-loop description 
and integrability properties. 
Sec.~(\ref{sec:Y}) is devoted to a brief summary of the relevant Y-system formulae.
In Sec.~(\ref{sec:results}), we present our results for the wrapping corrections. These are analyzed 
from the point of view of reciprocity in Sec.~(\ref{sec:reciprocity}), and in terms of a proposed BFKL
resummation in Sec.~(\ref{sec:bfkl}).

\section{Generalities on 3-gluon twist operators }
\label{sec:gluonic}

The 3-gluon operators are single-trace maximal helicity quasi-partonic operators which in the light-cone gauge take
the form 
\be
\label{eq:gaugeops}
\mathbb{O}_{N, 3}^A = \sum_{n_1+n_{2}+n_3=N} c_{n_1, n_{2}, n_3}\,\mbox{Tr}\left[
\partial_+^{n_1} \mathcal{A}(0)\,
\partial_+^{n_2} \mathcal{A}(0)\,
\partial_+^{n_3} \mathcal{A}(0)\right],
\ee
where $\mathcal A$ is the holomorphic combination of the physical gauge degrees of freedom $A^\mu_\perp$ and $\partial_+$ is the light-cone 
projected covariant derivative (in light-cone gauge the gauge links are absent).
The coefficients $\{c_{n_{1}, n_{2}, n_{3}}\}$ are such that $\mathbb{O}^A_{N, L}$ is an eigenvector of the dilatation
operator. The total Lorentz spin is $N$.

\medskip
The one-loop anomalous dimensions of the above operators can be found from the spectrum of a 
non-compact $XXX_{-3/2}$ spin chain with $3$ sites. 
At higher orders we are forced to abandon the quasipartonic detailed description and work in terms of superconformal multiplets. The identification of the $\mathfrak{psu}(2,2|4)$ primary of the multiplet where such operators appear as descendant can be done thanks to the work of ~\cite{Beisert:2004di}. We decompose
 the symmetric triple tensor product $(V_F\otimes V_F\otimes V_F)_S$
where $V_F$ is the singleton infinite dimensional irreducible representation of  $\mathfrak{psu}(2,2|4)$
\be
(V_F\otimes V_F\otimes V_F)_S = \mathop{\bigoplus_{n=0}}_{k\in\mathbb{Z}}^\infty c_n\left[V_{2k, n}+V_{2k+1, n+3}\right],
\ee
where $c_n$ are suitable multiplicities and $V_{n,m}$ well defined modules. For even $N$ and $m=2$, the one-loop lowest anomalous dimension in $V_{N,2}$ is associated with an unpaired state and has 
been proposed to be~\cite{Beisert:2004di} 
\be
\label{eq:oneloopmodule}
\gamma_{N,2} =  \frac{\lambda}{8\pi^2}\left[
2\,S_1\left(\frac{N}{2}+1\right)
+2\,S_1\left(\frac{N}{2}+2\right)+4\right]= \frac{\lambda}{8\pi^2}\left[
2\,S_1\left(\frac{N}{2}+1\right)+\frac{4}{N+4}+4\right],
\ee
where $g^2 = \lambda/(8\,\pi^2) = g_{\rm YM}^2\,N_c/(8\,\pi^2)$ is the scaled 't Hooft coupling, 
fixed in the planar $N_c\to\infty$ limit.
This result is in agreement with the analysis of maximal helicity 3 gluon operators in QCD~\cite{Belitsky:1999bf}.
The second term in (\ref{eq:oneloopmodule}) 
fully reveals the {\em violation of the maximum transcendentality principle}, as discussed (and exploited!)
in~\cite{Beccaria:2007pb}.

\medskip
These operators can also be presented as a special case of another family built with scalar fields~\cite{Freyhult:2009fc}.
They are the gauge theory dual of a minimal energy spinning 
string configuration with two spins, $S_1$ and $S_2$, in $AdS_5$ and charge $J$ in $S^5$. 
The field content of these operators can be schematically represented by
\be
\mbox{Tr}\left[ \mathcal{D}^{n+m}\,{\bar{\mathcal{D}}}^{m} \mathcal{Z}^{3}\right] \,.
\ee
The charges of the string are related to $m$ and $n$ through 
the identification
\be
S_1=n+m-\tfrac{1}{2}\,,\qquad S_2=m-\tfrac{1}{2}\,,\qquad J=L\,.
\ee

\subsection{Asymptotic anomalous dimension}

The asymptotic anomalous dimension has been studied at four loops in ~\cite{Beccaria:2007pb}. The three
loop results is 
\ba
\label{eq:gamma1}
\gamma_1 &=& 4\,S_1+\frac{2}{n+1}+4, \\
\label{eq:gamma2}
\gamma_2 &=& -2\,S_3-4\,S_1\,S_2-\frac{2\,S_2}{n+1}-\frac{2\,S_1}{(n+1)^2}-\frac{2}{(n+1)^3} + \\
&& -4\,S_2-\frac{2}{(n+1)^2}-8, \nonumber \\
\label{eq:gamma3}
\gamma_3 &=& 
+5\,S_5
+6\, S_2\, S_3
-4\, S_{2,3}
+4\, S_{4,1}
-8\,S_{3,1,1} \\
&&
+\left(4\, S_2^2+2\,S_4+8\,S_{3,1}\right)\,S_1 \nonumber \\
&&
+\frac{-S_4+4\,S_{2,2}+4\, S_{3,1}}{n+1}
+\frac{4\, S_1\, S_2+S_3}{(n+1)^2}
+\frac{2\, S_1^2+3\, S_2}{(n+1)^3} \nonumber \\
&&
+\frac{6\, S_1}{(n+1)^4}
+\frac{4}{(n+1)^5} \nonumber \\
&&
-2 \,S_4
+8\,S_{2,2} 
+8\, S_{3,1} \nonumber \\
&&
+\frac{4\, S_2}{(n+1)^2}
+\frac{4\, S_1}{(n+1)^3}
+\frac{6}{(n+1)^4} \nonumber \\ 
&&
+ 8\, S_2
+32, \nonumber
\ea
where $n = \frac{N}{2}+1$ and $S_{a, b, \dots}\equiv S_{a,b,\dots}(n)$ are nested harmonic sums. The four-loop result contains 
a contribution from the dressing phase and has a similar structure.
Many interesting structural properties as well as a reciprocity proof have been discussed in ~\cite{Beccaria:2007pb}. 
Our working hypothesis will be that the same structure is also shared by the leading wrapping correction,
precisely as it happens for genuinely $\mathfrak{sl}(2)$ operators.

\section{One-loop description of $\mathbb{O}_{N,3}^{\mathcal A}$ in $\mathfrak{sl}(2)$ grading}
\label{sec:oneloop}

We recall the one-loop description of the 3-gluon states under consideration. They are fully described in \cite{Beccaria:2007pb}. The 3-gluon states are associated with the following (higher) Dynkin diagram~\cite{Beisert:2005fw}  assignments
\be
\label{eq:dynkin}
\begin{minipage}{260pt}
\setlength{\unitlength}{1pt}
\small\thicklines
\begin{picture}(260,55)(-10,-30)
%
\put(-32,0){\line(1,0){22}}  
\put(  0,00){\circle{15}}
\put( -5,-5){\line(1, 1){10}}  
\put( -5, 5){\line(1,-1){10}}  
\dottedline{3}(8,0)(32,0)    
\put( 40,00){\circle{15}}     
\dottedline{3}(48,0)(72,0)   
\put( 80,00){\circle{15}}
\put( 75,-5){\line(1, 1){10}}  
\put( 75, 5){\line(1,-1){10}}  
\put( 80,-15){\makebox(0,0)[t]{$N+3$}}  
\put( 87,00){\line(1,0){26}} 
\put(120,00){\circle{15}}
\put(120,15){\makebox(0,0)[b]{$+1$}} 
\put(120,-15){\makebox(0,0)[t]{$N+4$}} 
\put(127,00){\line(1,0){26}} 
\put(160,00){\circle{15}}
\put(155,-5){\line(1, 1){10}}  
\put(155, 5){\line(1,-1){10}}  
\put(160,-15){\makebox(0,0)[t]{$N+2$}} 
\dottedline{3}(168,0)(192,0) 
\put(200,00){\circle{15}}
\put(200,-15){\makebox(0,0)[t]{$1$}} 
\dottedline{3}(208,0)(232,0) 
\put(240,00){\circle{15}}
\put(235,-5){\line(1, 1){10}} 
\put(235, 5){\line(1,-1){10}} 
\put(250,0){\line(1,0){20}} 
\end{picture}
\end{minipage}
\ee
As explained in \cite{Beccaria:2007pb}, after a chain of one-loop dualizations, 
the exact Bethe roots can be written in terms of those
of a $XXX_{-3/2}$ spin chain. The explicit roots can be given in terms
of the associated Baxter polynomials. Let us define the polynomial
\ba
P_{6}(u) &=& {}_{4}F_{3}\left(
\left.
\begin{array}{cc}
-\frac{N}{2}\quad \frac{N}{2}+4\quad \frac{1}{2}+i\,u\quad \frac{1}{2}-i\,u \\
2 \quad 2\quad 2
\end{array}
\right| 1\right).
\ea
In terms of $P_{6}$, we can write the Baxter polynomials $Q_{\ell}(u)\sim \prod_{i=1}^{K_{\ell}}(u-u_{\ell, i})$
for the roots at $\ell$-th node with $\ell=3,4,5$
\ba
Q_{5}(u) &=& (u+i)^{3}\,P_{6}\left(u+\frac{i}{2}\right)-(u-i)^{3}\,P_{6}\left(u-\frac{i}{2}\right), \\
Q_{4}(u) &=& \left(u+\frac{i}{2}\right)^{3}\,Q_{5}
\left(u+\frac{i}{2}\right)-\left(u-\frac{i}{2}\right)^{3}\,Q_{5}\left(u-\frac{i}{2}\right), \\
Q_{3}(u) &=& Q_{4}
\left(u+\frac{i}{2}\right)-Q_{4}\left(u-\frac{i}{2}\right).
\ea
The single root at node 6 is 
identically zero and $Q_{6}\sim u$. The Bethe equations satisfied by these polynomials are 
\ba
1 &=& \left. \frac{Q_{4}^{-}}{Q_{4}^{+}}\right|_{u_{3,p}}, \\
-\left(\frac{u_{4,p}+\frac{i}{2}}{u_{4,p}-\frac{i}{2}}\right)^{3} &=& \left. 
\frac{Q_{4}^{++}}{Q_{4}^{--}}
\frac{Q_{3}^{-}}{Q_{3}^{+}}
\frac{Q_{5}^{-}}{Q_{5}^{+}} \right|_{u_{4,p}}, \\
1 &=&  \left. \frac{Q_{4}^{-}}{Q_{4}^{+}}\frac{Q_{6}^{+}}{Q_{6}^{-}}\right|_{u_{5,p}}, \\
-1 &=&  \left. \frac{Q_{6}^{--}}{Q_{6}^{++}}\frac{Q_{5}^{+}}{Q_{5}^{-}}\right|_{u_{6,p}}.
\ea
For the following application, it will be convenient to dualize the Dynkin diagram and write it in $\mathfrak{sl}(2)$
grading. To this aim, we first  dualize at node 3. From 
\be
Q_{4}^{+}-Q_{4}^{-} = Q_{3}\,\widetilde Q_{3},
\ee
we deduce that $\widetilde Q_{3}$ is a constant, {\em i.e.} we don't have roots at node 3.
We then dualize at node 5. From 
\be
Q_{4}^{+}Q_{6}^{-}-Q_{4}^{-}Q_{6}^{+} = Q_{5}\,\widetilde Q_{5},
\ee
we deduce that the degree of $\widetilde Q_{5}$ is 2. Explicitly, after a short computation, one finds
\be
\widetilde Q_{5} = -\frac{1}{4}(N+4)^{2}-(N+3)(N+5)\,u^{2}.
\ee

\subsection{Summary}

After dualization (and omitting now the tilde), the one-loop Bethe equations in $\mathfrak{sl}(2)$ grading are associated with the diagram
\be
\label{eq:dynkin}
\begin{minipage}{260pt}
\setlength{\unitlength}{1pt}
\small\thicklines
\begin{picture}(260,55)(-10,-30)
\dottedline{3}(-32,0)(-10,0)  
\put(  0,00){\circle{15}}
\put( -5,-5){\line(1, 1){10}}  
\put( -5, 5){\line(1,-1){10}}  
\put(  7,00){\line(1,0){26}} 
\put( 40,00){\circle{15}}     
\put( 47,00){\line(1,0){26}} 
\put( 80,00){\circle{15}}
\put( 75,-5){\line(1, 1){10}}  
\put( 75, 5){\line(1,-1){10}}  
\dottedline{3}(88,0)(112,0)  
\put(120,00){\circle{15}}
\put(120,15){\makebox(0,0)[b]{$+1$}} 
\put(120,-15){\makebox(0,0)[t]{$N+4$}} 
\dottedline{3}(128,0)(152,0) 
\put(160,00){\circle{15}}
\put(155,-5){\line(1, 1){10}}  
\put(155, 5){\line(1,-1){10}}  
\put(160,-15){\makebox(0,0)[t]{$2$}} 
\put(167,00){\line(1,0){26}} 
\put(200,00){\circle{15}}
\put(200,-15){\makebox(0,0)[t]{$1$}} 
\put(207,00){\line(1,0){26}} 
\put(240,00){\circle{15}}
\put(235,-5){\line(1, 1){10}} 
\put(235, 5){\line(1,-1){10}} 
 \dottedline{3}(250,0)(270,0) 
%
\end{picture}
\end{minipage}
\ee
with the auxiliary polynomials
\ba
P_{6}(u) &=& {}_{4}F_{3}\left(
\left.
\begin{array}{cc}
-\frac{N}{2}\quad \frac{N}{2}+4\quad \frac{1}{2}+i\,u\quad \frac{1}{2}-i\,u \\
2 \quad 2\quad 2
\end{array}
\right| 1\right), \\
P_{5}(u) &=& (u+i)^{3}\,P_{6}^{+}-(u-i)^{3}\,P_{6}^{-}, 
\ea
and the Baxter polynomials
\ba
Q_{4}(u) &=& \left(u+\frac{i}{2}\right)^{3}\,P_{5}^{+}-\left(u-\frac{i}{2}\right)^{3}\,
P_{5}^{-}, \\
Q_{5}(u) &=& -\frac{1}{4}(N+4)^{2}-(N+3)(N+5)\,u^{2}, \\
Q_{6}(u) &=& u.
\ea
They obey the one-loop Bethe equations
\ba
&& -\left(\frac{u_{4,p}+\frac{i}{2}}{u_{4,p}-\frac{i}{2}}\right)^{3} = \left. 
\frac{Q_{4}^{--}}{Q_{4}^{++}}
\frac{Q_{5}^{+}}{Q_{5}^{-}} \right|_{u_{4,p}}, \\
&& 1 =  \left. \frac{Q_{4}^{+}}{Q_{4}^{-}}\frac{Q_{6}^{-}}{Q_{6}^{+}}\right|_{u_{5,p}}, \qquad
-1 =  \left. \frac{Q_{6}^{++}}{Q_{6}^{--}}\frac{Q_{5}^{-}}{Q_{5}^{+}}\right|_{u_{6,p}}.
\ea
This one-loop setup is enough to compute the leading wrapping correction as explained in the next Section.

\section{Y-system for the $\ads$ superstring}
\label{sec:Y}

\subsection{Generalities}

The Y-system is a set of functional equations for the functions $Y_{a,s}(u)$ defined on the fat-hook of
$\mathfrak{psu}(2,2|4)$~\cite{GKV}.
These equations are (their boundary conditions are discussed in \cite{Gromov:2009tv})
\be
\frac{Y^{+}_{a,s}\,Y^{-}_{a,s}}{Y_{a+1,s}\,Y_{a-1,s}}=\frac{
(1+Y_{a,s+1})(1+Y_{a,s-1})}{(1+Y_{a+1,s})(1+Y_{a-1,s})}.
\ee
The anomalous dimension of a generic state is given by the TBA formula 
\be
\label{eq:TBAenergy}
E = \sum_{i} \epsilon_{1}(u_{4,i})+\sum_{a\ge 1}\int_{\mathbb R}\frac{du}{2\pi i}
\frac{\partial \epsilon_{a}^{\star}}{\partial u}\,\log(1+Y^{\star}_{a,0}(u)).
\ee
In this formula, the dispersion relation is 
\be
\epsilon_{a}(u) = a+\frac{2\,i\,g}{x^{[a]}}-\frac{2\,i\,g}{x^{[-a]}},
\ee
and the star means evaluation in the mirror kinematics~\footnote{
We recall that the physical and mirror branches of the Zhukowsky relation
\be
x+\frac{1}{x} = \frac{u}{g},
\ee
are
\be
x_{\rm ph}(u) = \frac{1}{2}\left(\frac{u}{g}+\sqrt{\frac{u}{g}-2}\,\sqrt{\frac{u}{g}+2}\right), \qquad
x_{\rm mir}(u) = \frac{1}{2}\left(\frac{u}{g}+i\,\sqrt{4-\frac{u^{2}}{g^{2}}}\right).
\ee
Shifted quantities are defined as 
\be
F^{\underbrace{\pm\dots\pm}_{a}}(u) = F^{[\pm a]}(u) = F\left(u\pm i\,\frac{a}{2}\right).
\ee
For real $u$ and $a>0$, we have the relations
\be
\label{eq:mirror}
x_{\rm ph}^{[a]} = x_{\rm mir}^{[a]},\qquad
x_{\rm ph}^{[-a]} = 1/x_{\rm mir}^{[-a]}.
\ee
}. 
The Bethe roots $\{u_{4,i}\}$ are fixed by the exact Bethe equations (in physical kinematics)
\be
Y_{1,0}(u_{4}) = -1.
\ee
Any solution of the Y-system can be written as 
\be
Y_{a,s} = \frac{T_{a,s+1}\,T_{a,s-1}}{T_{a+1,s}T_{a-1,s}},
\ee
in terms of the solution $T_{a,s}$ of the Hirota integrable discrete equation
\be
T^{+}_{a,s}\,T^{-}_{a,s} = T_{a+1,s}T_{a-1,s}+T_{a,s+1}T_{a,s-1}.
\ee
This equation is gauge invariant ({\em i.e.} leads to the same $Y$) under
\be
T_{a,s}\la g_{1}^{[a+s]}\,g_{2}^{[a-s]}\,g_{3}^{[-a+s]}\,g_{4}^{[-a-s]}\,T_{a,s}.
\ee
The crucial assumption in the identification of relevant solutions to the Y-system is 
\be
Y^{\star}_{a\ge 1, 0}\sim \left(\frac{x^{[-a]}}{x^{[+a]}}\right)^{L},
\ee
for large $L$, or large $u$ (or small $g$). In this limit, it can be shown that the Hirota equation splits
in two $\mathfrak{su}(2|2)_{\rm L, R}$ wings. One can have a simultaneous finite large $L$ limit on both wings
after a suitable gauge transformation. The solution is then
\be
\label{eq:Ysolution}
Y_{a,0}(u) \simeq  \left(\frac{x^{[-a]}}{x^{[+a]}}\right)^{L}\,\frac{\phi^{[-a]}}{\phi^{[+a]}}\,
T^{\rm L}_{a,1}\,T^{\rm R}_{a,1},
\ee
where $\phi$ is an arbitrary function and $T^{\rm L, R}_{a, 1}$ are transfer matrices of the antisymmetric 
rectangular representations of $\mathfrak{su}(2|2)_{\rm L, R}$. They are given explicitly by the generating functional
\ba
\lefteqn{\sum_{a=0}^{\infty} (-1)^{a}\,T_{a,1}^{[1-a]}\,\ddb^{a} = \left(1-\frac{Q_{3}^{+}}{Q_{3}^{-}}\,\ddb\right)^{-1}\,
\left(1-\frac{Q_{3}^{+}}{Q_{3}^{-}}\frac{Q_{2}^{--}}{Q_{2}}\frac{R^{(+)-}}{R^{(-)-}}\,\ddb\right)} && \\
&& 
\qquad\times\left(1-\frac{Q_{2}^{++}}{Q_{2}}\frac{Q_{1}^{-}}{Q_{1}^{+}}\frac{R^{(+)-}}{R^{(-)-}}
\,\ddb\right)\,
\left(1-\frac{Q^{-}_{1}}{Q_{1}^{+}}\frac{B^{(+)+}}{B^{(-)+}}\frac{R^{(+)-}}{R^{(-)-}}\ddb\right)^{-1},
\nonumber
\ea
where $\ddb = e^{-i\partial_{u}}$ and 
\be
R_{}^{(\pm)} = \prod_{i=1}^{K_{4}} \frac{x(u)-x_{4,i}^{\mp}}{(x^{\mp}_{4,i})^{1/2}}, 
\qquad
B_{}^{(\pm)} = \prod_{i=1}^{K_{4}} \frac{\frac{1}{x(u)}-x_{4,i}^{\mp}}{(x^{\mp}_{4,i})^{1/2}}.
\ee
The function $\phi$ can be fixed as explained in \cite{GKV} and reads
\be
\frac{\phi^{-}}{\phi^{+}} = \sigma^{2}\,\frac{B^{(+)+}\,R^{(-)-}}{B^{(-)-}\,R^{(+)+}}\,
\frac{B^{+}_{1, \rm L}\,B^{-}_{3, \rm L}}{B^{-}_{1, \rm L}\,B^{+}_{3, \rm L}}\,
\frac{B^{+}_{1, \rm R}\,B^{-}_{3, \rm R}}{B^{-}_{1, \rm R}\,B^{+}_{3, \rm R}},
\ee
where $\sigma$ is the dressing phase~\footnote{By a suitable analytic continuation, it is possible to work in a strip
of the mirror kinematics where the dressing phase coincides with the physical one and gives higher order contributions 
that does not affect our computation.}. At weak coupling, 
evaluating the various terms at leading order in the mirror dynamics, the wrapping correction (second term
in the r.h.s. of  (\ref{eq:TBAenergy})) is simply given by the expression
\be
\Delta E = -\frac{1}{\pi}\sum_{a=1}^{\infty}\int_{\mathbb R}du\,Y_{a,0}^{\star}.
\ee

\subsection{Explicit formulae for the efficient computation of $Y^{\star}_{a, 0}$}

In the following, we shall need a compact efficient formula for the evaluation of  $Y^{\star}_{a, 0}$. 
According to (\ref{eq:Ysolution}), we need the contribution from the dispersion (ratio of $x^{\pm}$), 
the fusion of scalar factors ($\phi$ terms), and the $\mathfrak{su}(2|2)$ transfer matrices. 
The transfer matrices can be written in terms of the two spin dependent polynomials $Q_{4,5}$.
After a straightforward computation we obtain:

\medskip\noindent
\underline{\em Dispersion}

\medskip\noindent
This is the universal factor
\be
\left(\frac{4g^{2}}{a^{2}+4u^{2}}\right)^{3} .
\ee

\medskip\medskip\noindent
\underline{\em Fusion of scalar factors}

\medskip\noindent
After some manipulations, one finds  the formula
\be
\Phi^{\star}_{a} = [(Q^{+}_{4}(0)]^{2} \,\frac{Q_{4}^{[1-a]}}{Q_{4}^{[-1-a]}Q_{4}^{[a-1]}Q_{4}^{[a+1]}}
\,\frac{Q_{5}^{[-a]}}{Q_{5}}.
\ee

\medskip\noindent
\underline{\em{Left $su(2|2)$  wing} }

\medskip\noindent
The left wing has no excitations. It depends only on $Q_{4}$ and an efficient expression has been
already given in \cite{Beccaria:2010kd}
\be
T_{a,0}^{{\rm L}\ \star} = \left. i\,c\,g^{2}\,\frac{(-1)^{a+1}}{Q_{4}^{[1-a]}}
\mathop{\sum_{p=-a}^{a}}_{\Delta p = 2}\frac{Q_{4}^{[-1-p]}-Q_{4}^{[1-p]}}{u-\frac{i}{2}\,p} 
\right|_{Q_{4}^{[-a-1]} , Q_{4}^{[a+1]}\to 0},
\ee
where
\be
c = \sum_{j}\frac{1}{u_{4,j}+\frac{i}{2}}\frac{1}{u_{4,j}-\frac{i}{2}} = \left.
i\,(\log Q_{4})'\right|_{u=-i/2}^{u=+i/2},
\ee
is proportional to the one-loop anomalous dimension.

\medskip\medskip\noindent
\underline{\em{Right $su(2|2)$  wing} }

\medskip\noindent
The right wing depends  on $Q_{4}$ and $Q_{5}$. It also contains $Q_{6}$ which is simply $u$.
An efficient expression is
\be
T_{a,0}^{{\rm R}\ \star} = i\,(-1)^{a+1}\,\frac{Q_{5}^{[a]}}{Q_{4}^{[-(a-1)]}}\,
\mathop{\sum_{p=-(a-1)}^{a-1}}_{\Delta p=2}\frac{Q_{4}^{[p]}}{u+\frac{i}{2}\,p}
\left(
\frac{1}{Q_{5}^{[p+1]}}-\frac{1}{Q_{5}^{[p-1]}}
\right).
\ee

\section{Evaluation and results for the wrapping corrections}
\label{sec:results}

Let us inspect the first values of $Y^{\star}_{a,0}$ in order to identify the structure of the result. 
For all even $N$, we find empirically
that it is a rational function of $a$ and $u$. The expressions have a complexity which grows rapidly with $N$, but are 
very suitable for symbolic manipulations. For instance, at $N=2$ the explicit expression is 
\be
Y^{\star}_{a,0} \stackrel{N=2}{=} -2^{28}\cdot 5\cdot 7\,\,\frac{a^{2}\,g^{8}}{(a^{2}+4u^{2})^{4}}\,
\frac{N_{1}(a,u)\,N_{2}(a,u)}{D(a,u)\,D(-a,u)},
\ee
where
\ba
N_{1}(a,u) &=& \left(35 a^2-36\right) \left(a^4+44 a^2-288\right)-4 \left(35 a^4-3440 a^2+5328\right) u^2 \nonumber \\
&& -80 \left(35 a^2-272\right) u^4-6720 u^6, \\
N_{2}(a,u) &=& \left(7 a^6+120 a^4+944 a^2-1152\right)-4 \left(63 a^4+240 a^2-944\right) u^2 \nonumber \\
&& -80 \left(7 a^2+72\right) u^4+2240 u^6, \\
D(a,u) &=& \left(35 a^6+210 a^5+920 a^4+2280 a^3+3536 a^2+3072 a+1152\right)^2 +  \\
&& +8 \left(3675 a^{10}+36750 a^9+206850 a^8+772800 a^7+2155360 a^6+4582560 a^5+\nonumber \right. \\
&& \left. +7220320
   a^4+8052480 a^3+6144256 a^2+2982912 a+645120\right) u^2+\nonumber \\
   && +16 \left(18375 a^8+147000 a^7+569800 a^6+1360800 a^5+2603200 a^4+4084800 a^3+\right. \nonumber \\
   && \left. +3847360 a^2+1585920
   a+614656\right) u^4+1280 \left(1225 a^6+7350 a^5+15610 a^4+\right. \nonumber \\
   && \left. +13440 a^3+24016 a^2+40752 a-4912\right) u^6+3840 \left(1225 a^4+4900 a^3+\right. \nonumber \\
   && \left. +1820 a^2-6160
   a+9088\right) u^8+358400 \left(21 a^2+42 a-58\right) u^{10}+5017600 u^{12}.\nonumber
   \ea
The wrapping correction is evaluated as usual at leading order as 
\be
W_{N=2} = -\frac{1}{\pi}\sum_{a\ge 1}\int_{\mathbb R} Y^{\star}_{a,0} (u) du.
\ee
We checked that this can be obtained by summing the residues in $u=i\frac{a}{2}$. 
Notice also that the residue is a rational function of $a$ of the form 
\be
\mathcal{R}(a) = \mbox{Res}_{u=ia/2} Y^{\star}_{a,0}(u) = \frac{P(a)}{a^{5}\,\left[
Q_{4}^{+}(i\,a)\,
Q_{4}^{-}(i\,a)\,
\right]^{4}},
\ee
where $P$ is a polynomial.
The sum over $a$ can be done exploiting the following important properties. First, the singular part 
of $\mathcal{R}(a) $
at $a=0$ is a combination of two poles $\sim a^{-3}$ and $a^{-5}$. These contributions can be summed giving terms proportional
to $\zeta_{3}$ and $\zeta_{5}$. Then, the regularized rational function 
\be
\widetilde {\mathcal{R}}(a) = \mathcal{R}(a) -\mbox{``pole part at $a=0$``},
\ee
obeys the remarkable property
\be
\sum_{a=1}^{p}\widetilde {\mathcal{R}}(a) = \frac{\widetilde P(p)}{\left[Q_{4}^{+}(i\,p)\right]^{4}},
\ee
where $\widetilde P$ is another polynomial with the same degree of the denominator. It can be determined easily by polynomial interpolation. Then, the infinite sum is just the trivial $p\to \infty$ limit.

\medskip
Repeating the calculation for higher values of $N$, we find the following list  (omitting $g^{8}$)
\ba
W_{N=2} &=& \frac{5348840}{2187}+\frac{528640}{81}\,\zeta_3-\frac{89600}{9}\,\zeta_5 \nonumber \\ 
W_{N=4} &=& \frac{1216307603}{331776}+\frac{235081}{24}\,\zeta_3-14910\,\zeta_5 \nonumber \\ 
W_{N=6} &=& \frac{4018092206843}{810000000}+\frac{24863234}{1875}\,\zeta_3-\frac{100848}{5}\,\zeta_5 \nonumber \\ 
W_{N=8} &=& \frac{27624795456401}{4374000000}+\frac{56992078}{3375}\,\zeta_3-\frac{231088}{9}\,\zeta_5 \nonumber \\ 
W_{N=10} &=& \frac{206037054943950841}{26682793200000}+\frac{3718258258}{180075}\,\zeta_3-\frac{1538160}{49}\,\zeta_5 \nonumber \\ 
W_{N=12} &=& \frac{143200464784761259303}{15613245849600000}+\frac{80764059527}{3292800}\,\zeta_3-\frac{521985}{14}\,\zeta_5 \nonumber \\ 
W_{N=14} &=& \frac{767836561931733099146293}{72027074544768000000}+\frac{32079436339187}{1125211500}\,\zeta_3-\frac{24571256}{567}\,\zeta_5 \nonumber \\ 
W_{N=16} &=& \frac{1857233391274028370487}{152438253004800000}+\frac{77595407257}{2381400}\,\zeta_3-\frac{148580}{3}\,\zeta_5 \nonumber \\ 
W_{N=18} &=& \frac{102022930034541927436784638469}{7426475758114503552000000}+\frac{640511693446157}{17433038700}\,\zeta_3-\frac{20271464}{363}\,\zeta_5. \nonumber
\ea
We extended the list up to $N=70$. The general form is always
\be
W_{N} = \left(r_{0, N}+r_{3,N}\,\zeta_3+r_{5,N}\,\zeta_5\right)\,g^{8},
\ee
with rational coefficients $r_{0, N}$, $r_{3,N}$, and $r_{5,N}$.

\subsection{Closed fomulae}

Given our long list of explicit values $\{W_{N}\}$, we can look for a closed formula based on the 
structure of the asymptotic anomalous dimensions. Notice also that
from the expression of $T^{{\rm L}\ \star}$, we see that the wrapping correction is proportional to the one-loop
anomalous dimension, {\em i.e.} the combination 
\be
4\,S_{1}(n)+\frac{2}{n+1}+4,\qquad n=\frac{N}{2}+1.
\ee

\subsubsection*{Coefficient of $\zeta_{5}$}

After some trial and error we find $S_{a,b,\dots}\equiv S_{a,b,\dots}(n)$
\be
r_{5,N} = 80\, \left(4\,S_{1}+\frac{2}{n+1}+4\right)\,\left(
-4(n+1)+\frac{1}{n+1}
\right).
\ee

\subsubsection*{Coefficient of $\zeta_{3}$}

Inspired by the structure of the previous coefficient we find

\ba
\lefteqn{r_{3,N} = 16\, \left(4\,S_{1}+\frac{2}{n+1}+4\right)\times} && \\
&& \,\left[
8\,(n+1)\,S_{2}+8+ \frac{2}{n+1}(2-S_{2}) -\frac{2}{(n+1)^{2}}-\frac{1}{(n+1)^{3}}
\right].\nonumber
\ea

\subsubsection*{Purely rational part}

With major effort, we obtain 

\ba
\lefteqn{r_{0,N} = 2\, \left(4\,S_{1}+\frac{2}{n+1}+4\right)\times} && \\
&& \,\left[
16\,(n+1)\,(2\,S_{2,3}-S_{5})+32\,S_{3}+\frac{4}{n+1}(S_{5}-2S_{2,3}+4S_{3})+ \right.\nonumber\\
&& \left. +\frac{8}{(n+1)^{2}}(-S_{3}+2)+\frac{4}{(n+1)^{3}}(-S_{3}+4)-\frac{4}{(n+1)^{5}}-\frac{1}{(n+1)^{6}}
\right].\nonumber
\ea

\section{Large $N$ expansion and generalized Gribov-Lipatov reciprocity}
\label{sec:reciprocity}

Let us factor out the one-loop anomalous dimension and write ($k=0,3,5$)
\be
r_{k, N} = \left(4\,S_{1}+\frac{2}{n+1}+4\right)\,\widetilde r_{k, N}.
\ee
By standard methods, we worked out the large $N$ of $\widetilde r_{k,N}$. It reads
\ba
\widetilde r_{5, N} &=& -320\, n-320+\frac{80}{n}-\frac{80}{n^2}+\frac{80}{n^3}-\frac{80}{n^4}+\frac{80}{n^5}+\dots, \\
\widetilde r_{3, N} &=& 128\, \zeta _2\, n+128 \zeta _2-\frac{32 \zeta _2}{n}+\frac{16 \left(6 \zeta _2-4\right)}{3 n^2}-\frac{16 \left(6 \zeta _2-8\right)}{3 n^3}+\nonumber \\
&& +\frac{32 \zeta
   _2-\frac{272}{5}}{n^4}+\frac{\frac{704}{15}-32 \zeta _2}{n^5}+\dots, \\
   \widetilde r_{0, N} &=& \left(320 \zeta _5-128 \zeta _2 \zeta _3\right) n+\left(320 \zeta _5-128 \zeta _2 \zeta _3\right)+\frac{32 \zeta _2 \zeta _3-80 \zeta _5}{n}+ \\
   && -\frac{16 \left(6
   \zeta _2 \zeta _3+2 \zeta _3-15 \zeta _5-2\right)}{3 n^2}+\frac{16 \left(6 \zeta _2 \zeta _3+4 \zeta _3-15 \zeta _5-4\right)}{3 n^3}+\nonumber \\
   && -\frac{8 \left(60 \zeta
   _2 \zeta _3+51 \zeta _3-150 \zeta _5-36\right)}{15 n^4}+\frac{16 \left(30 \zeta _2 \zeta _3+22 \zeta _3-75 \zeta _5+8\right)}{15
   n^5}+\dots.\nonumber
   \ea
Remarkably, the leading $\mathcal{O}(n)$ and next-to-leading $\mathcal{O}(n^{0})$ contributions 
cancel. The resulting total wrapping contribution is thus (including the one-loop factor, and writing the expansion
in the $N$ variable)
\ba
\frac{1}{g^{8}}\,W_{N} &=& -\frac{512}{3} \left(3 \zeta _3-1\right) \,\frac{\log\overline N+1}{N^{2}}
+\frac{2048}{3} \left(3 \zeta _3-1\right) \frac{2 \log\overline N+1}{N^{3}} + \nonumber \\
&& -\frac{1536}{5} \left(77 \zeta _3-24\right)\,\frac{\log\overline N}{N^{4}} 
-\frac{512}{45} \left(264 \zeta _3-43\right)\,\frac{1}{N^{4}} + \nonumber \\
&& \frac{8192}{15} \left(213 \zeta _3-56\right)\,\frac{\log\overline N}{N^{5}} 
-\frac{2048}{45} \left(543 \zeta _3-316\right)\,\frac{1}{N^{5}}+\dots. 
\ea
where $\overline N = \frac{N}{2}\,e^{\gamma_{\rm E}}$. Notice that all terms containing $\zeta_{2} = \pi^{2}/6$
as well as $\zeta_{5}$ cancel.

\medskip
As an important remark, we emphasize that the leading wrapping correction is subleading at large $N$ compared with the asymptotic contribution and does
not change the scaling function already computed for these operators in \cite{Beccaria:2007pb} and 
in more general cases in \cite{Freyhult:2009fc}.

\subsection{Generalized Gribov-Lipatov reciprocity}

As discussed in \cite{Beccaria:2007pb}, the asymptotic anomalous dimensions can be expanded at large $J$
where $J^{2} = n\,(n+2)$ and turn out to admit an expansion in even powers of $1/J$. The absence of odd powers
is the so-called reciprocity property~\footnote{This means that the large $N$ expansion can be organized in a part with even powers of $1/N$ and an induced part with odd powers of $1/N$ completely determined by the former.}
For instance, the one-loop factor has the expansion
\ba
4\,S_{1}(n)+\frac{2}{n+1}+4 &=& 
(2 \log \overline J^{2}+4)+\frac{5}{3 J^{2}}-\frac{19}{30 J^{4}}+\frac{79}{315 J^{6}}-\frac{1}{420 J^{8}}+\dots.
\ea
where $\overline J = J e^{\gamma_{\rm E}}$.

We now look at the reciprocity property of the wrapping correction associated with the $\widetilde r$ factors.
From the previous expansions we find 
\ba
\widetilde r_{5, N} &=& J\,\left(-320-\frac{80}{J^{2}}+\frac{10}{J^{6}}-\frac{25}{2 J^{8}}+\frac{105}{8
J^{10}}+\dots\right), \\
\widetilde r_{3, N} &=&  128 \zeta _2 J+\frac{32 \zeta _2}{J}-\frac{64}{3 J^2}+\frac{464}{15 J^4}-\frac{4 \zeta _2}{J^5}-\frac{4688}{105 J^6}+\frac{5 \zeta _2}{J^7}+\frac{7088}{105
   J^8}+\dots, \\
\widetilde r_{0, N} &=&    \left(320 \zeta _5-128 \zeta _2 \zeta _3\right) J+\frac{80 \zeta _5-32 \zeta _2 \zeta _3}{J}-\frac{32 \left(\zeta _3-1\right)}{3 J^2}+\frac{\frac{232 \zeta
   _3}{15}-\frac{352}{15}}{J^4}+\frac{4 \zeta _2 \zeta _3-10 \zeta _5}{J^5}+\nonumber \\
   && +\frac{\frac{4834}{105}-\frac{2344 \zeta _3}{105}}{J^6}+\frac{\frac{25 \zeta
   _5}{2}-5 \zeta _2 \zeta _3}{J^7}+\frac{\frac{3544 \zeta _3}{105}-\frac{83956}{945}}{J^8}+\dots.
   \ea
Notice that none of the three terms is separately reciprocity respecting. However, 
the combination appearing in the wrapping correction reads
\ba
\lefteqn{\zeta_{5}\,\widetilde r_{5, N} +\zeta_{3}\,\widetilde r_{3, N}+\widetilde r_{0, N} = } && \\
&& \frac{\frac{32}{3}-32 \zeta _3}{J^2}+\frac{\frac{232 \zeta _3}{5}-\frac{352}{15}}{J^4}+\frac{\frac{4834}{105}-\frac{2344 \zeta _3}{35}}{J^6}+\frac{\frac{3544
   \zeta _3}{35}-\frac{83956}{945}}{J^8}+\frac{\frac{271768}{1485}-\frac{9512 \zeta _3}{55}}{J^{10}}+\nonumber \\
   && \frac{\frac{1872392 \zeta
   _3}{5005}-\frac{20053258}{45045}}{J^{12}}+\frac{\frac{87933002}{61425}-\frac{524872 \zeta _3}{455}}{J^{14}}+\frac{\frac{4917304 \zeta
   _3}{935}-\frac{5747755528}{883575}}{J^{16}}+\dots. \nonumber
\ea
All the odd powers of $1/J$ cancel proving that the reciprocity property does hold.

\section{BFKL analytic continuation}
\label{sec:bfkl}

In the $\mathfrak{sl}(2)$ sector, and for length $L=2$, it is possible to explore the analytic continuation of the anomalous dimensions at negative spin. The leading and next-to-leading poles are captured by the BFKL equation
thus providing a strong cross check of the calculation. For the 3-gluon states, we shall now compute the BFKL poles 
and show that a very simple and natural modification of the twist-2 BFKL equation predicts the correct pole 
structure~\footnote{This section has been prepared together with Nikolay Gromov. We thanks him for his 
kind help and insight.}.

\subsection{Continuation at $n=-1$}

The four loop anomalous dimension is 
\ba
\gamma(g) &=& g^{2}\,\gamma_{1}^{\rm ABA}+g^{4}\,\gamma_{2}^{\rm ABA}+g^{6}\,\gamma_{3}^{\rm ABA}
+\\
&& + g^{8}\,\left(
\gamma_{4}^{\rm ABA}+\zeta_{3}\,\gamma_{4}^{\rm Dressing}+W
\right)+\dots.\nonumber
\ea
Since we have the explicit closed form of all terms as functions of $N$, we analytically continue in the 
variable $n = \frac{N}{2}+1$ around $n=-1$. Setting $n=-1+\omega$, we find~\footnote{Notice a factor 2
of difference in the definition of $g^{2}$ in this paper  and in \cite{Beccaria:2007pb}.}
\ba
\gamma_{1}^{\rm ABA} &=& -\frac{4}{\omega }+\dots,  \\
\gamma_{2}^{\rm ABA} &=& \frac{8}{\omega ^2}+\frac{4 \pi ^2}{3 \omega }+\dots, \\
\gamma_{3}^{\rm ABA} &=&  \frac{\bf 0}{\omega^{3}}-\frac{16 \left(-3 \zeta _3+\pi ^2+12\right)}{3 \omega ^2}+\dots,\\
\gamma_{4}^{\rm ABA} &=& -\frac{4}{\omega ^7}+\frac{24}{\omega ^6}+\frac{4 \left(\pi ^2-24\right)}{3 \omega ^5}-\frac{8 \left(3 \zeta _3+2 \pi ^2+36\right)}{3
   \omega ^4}+\dots, \\
\gamma_{4}^{\rm Dressing} &=& -\frac{32}{\omega ^4}+\dots.
   \ea
   The expansion of the wrapping contribution is 
   \ba
W &=& \frac{4}{\omega ^7}-\frac{24}{\omega ^6}+\frac{32-\frac{4 \pi ^2}{3}}{\omega ^5}+\frac{-24 \zeta _3+\frac{16 \pi ^2}{3}+64}{\omega
   ^4}+\dots. 
\ea
Summing all terms for the $g^{8}$ contribution, we get
\be
\gamma_{4} = \frac{\bf 0}{\omega^{7}}+\frac{\bf 0}{\omega^{6}}+\frac{\bf 0}{\omega^{5}}+\frac{-64 \zeta _3-32}{\omega ^4}+\frac{160 \zeta _3}{\omega ^3}+\dots.
\ee
The cancellation of the three leading poles (present in $\gamma_{4}^{\rm ABA}$ and canceling
against $W$) is remarkable. We can reproduce the leading poles from a BFKL-like equation as in \cite{Kotikov:2007cy,Bajnok:2008qj}
as follows. In the Konishi case, one defines 
\be
\chi_{a}(z) = \psi(-z)+\psi(z+1+a)-2\psi(1),
\ee
and solves perturbatively the equation
\be
\label{eq:bfkl-konishi}
\frac{\omega}{-g^{2}} = \chi_{0}\left(\frac{\gamma}{2}\right),
\ee
where $\gamma$ is written as a power series in $g^{2}$. 
To extend this equation to our case, we notice that the BFKL kernel can be written as
\be
\chi_{0}(z) = h(z)+h(-z-1),
\ee
where $h(z) = \psi(z+1)-\psi(1) \equiv S_{1}(z)$ is the analytic continuation of the basic harmonic sum, proportional
to the one-loop anomalous dimension. 
In our case, the one-loop anomalous dimension (\ref{eq:gamma1}) can be 
written in the form 
\be
\gamma_{1}(n) = 2\,S_{1}(n)+2\,S_{1}(n+1)+4.
\ee
We thus led to replace in (\ref{eq:bfkl-konishi})
\be
\chi_{0}(z)\la \chi_{0}(z)+\chi_{0}(z+1) \equiv 2\chi_{1}(z).
\ee
Thus, we have to solve
\be
\label{eq:bfkl}
\frac{\omega}{-g^{2}} = 2\,\chi_{1}\left(\frac{\gamma}{2}\right).
\ee
Expanding at weak coupling, we indeed obtain
\ba
\gamma &=& \left(-\frac{4}{\omega}+\dots\right)\,g^{2}+
\left(\frac{8}{\omega^{2}}+\dots\right)\,g^{4}+
\left(\frac{\mathbf{0}}{\omega^{3}}+\dots\right)\,g^{6}+\\
&&+
\left(-\frac{32\,(1+2\,\zeta_{3})}{\omega^{4}}+\dots\right)\,g^{8}
+
\left(\frac{512\,\zeta_{3}}{\omega^{5}}+\dots\right)\,g^{10}+\dots.
\ea
This is in full agreement with our four loop results. 
Notice that the agreement of the four loop term is achieved thanks to the non-trivial wrapping contribution.
The higher order poles are a prediction.

\medskip
One can also attempt to write down a NLO BFKL equation of the form 
\be
\frac{\omega}{-g^{2}} = 2\,\chi_{1}\left(\frac{\gamma}{2}\right)-g^{2}\,\delta(\gamma),
\ee
and fix the expansion of $\delta(\gamma)$ from the coefficients of the next to leading poles. At the four loop 
order one finds the remarkably simple result
\be
\delta(\gamma) = \frac{2}{\gamma}\left[
4\,\zeta_{2}+2\,(4+\zeta_{2}-\zeta_{3})\,\gamma-(8+\zeta_{2}-7\,\zeta_{3})\,\gamma^{2}+\dots
\right]
\ee
with all integer coefficients.

\subsection{Continuation at $n=-2$}

We can also compute the singular expansion around $n=-2$. We do this with the aim of comparing with what happens
in the Konishi case \cite{Kotikov:2007cy,Bajnok:2008qj}. Setting $n=-2+\omega$, and considering only the leading poles,  the result is 
\ba
\gamma_{1}^{\rm ABA} &=& -\frac{8}{\omega }+\dots,  \\
\gamma_{2}^{\rm ABA} &=& -\frac{8}{\omega ^3}+\dots, \\
\gamma_{3}^{\rm ABA} &=& -\frac{8}{\omega ^5}+\dots,\\
\gamma_{4}^{\rm ABA} &=& -\frac{8}{\omega ^7}+\dots, \\
\gamma_{4}^{\rm Dressing} &=& -\frac{128}{\omega ^4}+\dots.
   \ea
   The expansion of the wrapping contribution is 
   \ba
W &=& \frac{96}{\omega ^6}+\dots. 
\ea
We see the same pattern as in the Konishi case. In particular, the ABA leading poles are of the same order
as in Konishi, {\em i.e.} $\sim g^{2n}/\omega^{2n-1}$, and wrapping correction does not change them.

\section{Conclusions}

In this paper we have applied the Y-system formalism proposed in \cite{GKV} to the computation of the 
wrapping correction for a class of twist operators reducing at one-loop to 3-gluon maximal helicity quasipartonic operators. We provided the leading four-loop 
contribution as a closed function 
of the operator spin $N$ by a generalized transcendentality Ansatz already found in the asymptotic contribution.
The result can be checked by means of two important physical constraints: A generalized reciprocity
of the large $N$ expansion, and a BFKL-like resummation of the leading pomeron poles. Both tests are passed
extending similar conclusions holding for the simpler $\mathfrak{sl}(2)$ twist operators. 

Our analysis is a novel test of the Y-system for the $\ads$ superstring involving a larger part of the 
$\mathfrak{psu}(2,2|4)$ Dynkin diagram and leading to exact prediction for a new set of short operators.
It would be very interesting to extend the analysis to the strong coupling regime by solving 
the full (numerical or possibly semiclassical) TBA equations. In this perspective, it would be very nice to 
assess the validity and consequences of the reciprocity and BFKL constraints for the string duals of the considered
operators.

\section*{Acknowledgments}

We thank  Nikolay Gromov and  Fedor Levkovich-Maslyuk  for helpful 
and stimulating discussions.

\end{document}